\documentclass[12pt]{article}
\usepackage{amsfonts,amssymb,amsmath,amsthm,amscd,bbm,enumitem,tcolorbox}
\usepackage[mathscr]{euscript} %use for \N for Normal distribution, closer to the original mathcal
\usepackage{natbib}
\usepackage[onehalfspacing]{setspace}
\usepackage{color,graphicx,caption,subcaption,hyperref}
\hypersetup{pdfborder = {0 0 0},colorlinks=true,linkcolor=blue,urlcolor=blue,citecolor=blue}
\usepackage[margin=1in]{geometry}
\usepackage[normalem]{ulem}

\newcommand\inde{\protect\mathpalette{\protect\independenT}{\perp}}
\def\independenT#1#2{\mathrel{\rlap{$#1#2$}\mkern2mu{#1#2}}}

\newcommand{\forloop}[5][1]{\setcounter{#2}{#3}\ifthenelse{#4}{#5\addtocounter{#2}{#1}\forloop[#1]{#2}{\value{#2}}{#4}{#5}}{}}

\newcommand{\I}{\mathbbm{1}}

\newcommand{\E}{\mathbb{E}}

\renewcommand{\P}{\mathbb{P}}

\begin{document}

\title{Covariate Adjustment in Regression Discontinuity Designs\thanks{We thank our collaborators Sebastian Calonico, Max Farrell, Michael Jansson, Xinwei Ma, Gonzalo Vazquez-Bare and Jose Zubizarreta for stimulating discussions on the topic of this chapter. Cattaneo and Titiunik gratefully acknowledges financial support from the National Science Foundation (SES-2019432), and Cattaneo gratefully acknowledges financial support from the National Institutes of Health (R01 GM072611-16).
}}

\author{
	Matias D. Cattaneo\footnote{Department of Operations Research and Financial Engineering, Princeton University.}\and
	Luke Keele\footnote{Department of Surgery and Biostatistics, University of Pennsylvania.}\and
	Roc\'{i}o Titiunik\footnote{Department of Politics, Princeton University.}}

\maketitle
\setcounter{page}{0}\thispagestyle{empty}

\begin{abstract}
    The Regression Discontinuity (RD) design is a widely used non-experimental method for causal inference and program evaluation. While its canonical formulation only requires a score and an outcome variable, it is common in empirical work to encounter RD analyses where additional variables are used for adjustment. This practice has led to misconceptions about the role of covariate adjustment in RD analysis, from both methodological and empirical perspectives. In this chapter, we review the different roles of covariate adjustment in RD designs, and offer methodological guidance for its correct use.   
\end{abstract}


\newpage

\setcounter{tocdepth}{2}

\tableofcontents

\setcounter{page}{0}
\thispagestyle{empty}

\newpage

\maketitle
\newpage

\doublespacing

\section{Introduction}

In causal inference and program evaluation, a central goal is to learn about the causal effect of a policy or treatment on an outcome of interest. There are a variety of methodological approaches for the identification, estimation, and inference for causal treatment effects, depending on the specific application. In experimental settings, where the treatment assignment mechanism is known, studying treatment effects is relatively straightforward because the methods are based on assumptions that are known to be true by virtue of the treatment assignment rule. In contrast, in observational studies, the assignment mechanism is unknown and researchers are forced to invoke identifying assumptions whose validity is not guaranteed by design. Well known observational study methods include selection on observables, instrumental variables, difference-in-differences, synthetic controls, and regression discontinuity designs.\footnote{See, for example, \citet{Rosenbaum2010-book}, \citet{ImbensRubin2015-book} and \citet{Abadie-Cattaneo_2018_ARE} for reviews on causal inference and program evaluation, and \citet{Titiunik_2021_HandbookCh} for further discussion on the role of natural experiments in observational research.}

In experimental and observational studies, baseline covariates generally serve different purposes. In experimental settings, covariate adjustment based on pre-intervention measures is often used either for efficiency gains or for the evaluation of treatment effect heterogeneity. On the other hand, in many observational studies, the primary purpose of covariate adjustment is for the identification of causal effects. For example, under selection on observables, the researcher posits that, after conditioning on a set of pre-intervention covariates, the assignment mechanism is not a function of potential outcomes, thereby mimicking an experimental setting after conditioning on those covariates. Under this assumption, researchers can obtain valid estimates of treatment effects by comparing groups of units whose covariate values are similar. Implementation of covariate adjustment methods in experimental and observational studies can be done in different ways, including via regression methods, inverse probability weighting, and matching. However, the validity of covariate adjustment methods is not automatically guaranteed because it requires additional (usually stronger) assumptions, if valid at all. This is particularly true in the case of the regression discontinuity (RD) design.

Covariate adjustment in RD designs has been employed in a variety of ways and for different purposes in recent years. Because the RD design has a unique feature relative to other non-experimental research strategies, covariate adjustment has distinctive implications for the analysis and interpretation of RD treatment effects. This has led to confusions and misconceptions about the role of auxiliary covariates in RD designs. The main purpose of this chapter is to review the different roles of covariate adjustment in the RD literature, and to offer methodological guidance for its correct use in applications.

To streamline the presentation, we do not provide a comprehensive review of RD designs and methods.\footnote{For recent introductions and practical guides see, for example, \citet*{Cattaneo-Titiunik-VazquezBare_2020_BookCh}, \citet*{Cattaneo-Idrobo-Titiunik_2020_CUP,Cattaneo-Idrobo-Titiunik_2022_CUP}, \citet*{Cattaneo-Keele-Titiunik_2022_MedReview}, \citet*{Cattaneo-Titiunik_2022_ARE}, and references therein.} We focus our discussion on the canonical (sharp) RD design, where all units receive an observable score and a treatment is assigned to all units whose value of the score exceeds a known and fixed cutoff, and withheld from units whose score is below it. Under appropriate conditions, the abrupt change in the probability of receiving treatment that occurs at the cutoff justifies comparing units just above the cutoff to units just below it to learn about a causal effect of the treatment. For this canonical RD setup, and employing the taxonomy introduced by \citet*{Cattaneo-Titiunik-VazquezBare_2017_JPAM}, we discuss the different roles of covariate adjustment within the two leading conceptual frameworks for the interpretation and analysis of RD designs: the continuity framework and the local randomization framework.

One of the most important roles of baseline covariates in the canonical RD designs is for falsification or validation purposes. In the continuity framework for RD analysis, covariates are used to verify that the effect of the treatment on the covariates at the cutoff is zero, offering evidence in support of the assumption that only the probability of treatment assignment/status given the score changes discontinuously at the cutoff. In the local randomization approach, covariates are often used to select the window around the cutoff where treated and control units are similar to each other, and where the assumption of as-if random treatment assignment is assumed to hold.

The canonical RD setup is sometimes augmented with additional covariate information for other purposes beside falsification testing, and the role of these additional covariates varies depending on the setting and specific goals. For example, additional covariates may be used for efficiency gains or for the analysis of treatment effect heterogeneity. This use of pre-intervention covariates is motivated by the analogy between randomized experiments and local randomization in RD designs. Alternatively, covariates may be used to construct other treatment effects of interest or for extrapolation purposes (i.e., external validity). This occurs, for instance, when the additional covariates capture other features of the design such as geographic location, evolution over time, or group identity. Finally, sometimes covariates are also used in conjunction with statistical adjustment with the hope of ``fixing'' an otherwise invalid RD design. See \citet*{Cattaneo-Titiunik_2022_ARE} for a comprehensive review and references.

In this chapter, we discuss the most common uses that have been proposed for covariate adjustment in the RD literature, including efficiency, heterogeneity, and extrapolation. In each case, we discuss whether these methods are best justified from a continuity or a local randomization perspective, and how they methodologically relate to each other. Our main focus is conceptual, and mostly from an identification perspective, although we do discuss implementation issues when appropriate. Covariate adjustment in RD designs can be implemented by either local linear regression, inverse probability weighting, or ``matching'' methods more generally.

The rest of the chapter is organized as follows. The next section reviews the different roles of covariate adjustment in RD designs. Then, in Section \ref{sec: Can Covariates Fix a Broken RD Design?}, we discuss the important question of whether the use of covariate adjustment (e.g., adding fixed effects for group of units) can ``restore'' the validity of an RD design where the key identifying assumptions do not hold (e.g., settings where units are suspected to have sorted around the cutoff). Section \ref{sec: Recommendations for Practice} offers generic recommendations for practice, and Section \ref{sec: Conclusion} concludes.

\section{Covariate Adjustment in RD Designs}\label{sec: Covariate Adjustment in RD Designs}

For each type of covariate adjustment in RD designs, we discuss specific implications of using covariates. In particular, we outline how covariate adjustment can lead to unintended redefinition of the estimand to a quantity that does not coincide with a parameter of interest. We clarify whether the covariate adjustment applies to the continuity-based, the local randomization framework, or both. While most of the available proposals were introduced within one of these two conceptual frameworks, we also discuss the potential implications that each of the covariate adjustment approaches may have for the other. In this discussion it is important to remember that the core identifying assumptions differ between the two frameworks, and consequently so does the RD parameter of interest. For simplicity, we focus only on sharp RD designs (i.e., settings with perfect treatment compliance or intention-to-treat analysis), but the references given often include discussions of fuzzy, kink, and other related RD setups. 

\subsection{Overview of the Canonical RD Design}

We assume that the observed data is $(Y_i,D_i,X_i,Z_i)$, $i=1,2,\dots,n$, where $Y_i=Y_i(0)(1-D_i)+Y_i(1)D_i$, $Y_i(0)$ is the potential outcome under control, $Y_i(1)$ is the potential outcome under treatment, $D_i=\I(X_i\geq c)$ denotes the treatment assignment rule, $X_i$ denotes the score variable, $c$ denotes the cutoff, and $Z_i$ denotes other pre-determined covariates. Depending on the framework considered, the potential outcomes are taken to be random or non-random, and the data can be seen either as a (random) sample from some large population or as a finite population. The core assumption in the RD design is that the probability of treatment assignment changes abruptly at the cutoff from zero to one, that is, $\P[D_i=1|X_i=x] = 0$ if $x<c$ and $\P[D_i=1|X_i=x] = 1$ if $x\geq c$.

In the continuity framework, the data is assumed to be a random sample and the potential outcomes are assumed to be random variables. In this context, the most common parameter of interest is the average treatment effect at the cutoff,
\[\tau_\texttt{SRD} = \E[Y_i(1)-Y_i(0)|X_i=c],\]
and the key identifying assumption is the continuity of the conditional expectations of the potential outcomes given the score: $\E[Y_i(1) \;|\; X_i=x]$ and $\E[Y_i(0) \;|\; X_i=x]$ are continuous in $x$ at $c$, which leads to the well-known continuity-based identification result \citep*{Hahn-Todd-vanderKlaauw_2001_ECMA}:
\[ \tau_\texttt{SRD} = \lim_{x\downarrow{c}} \E[Y_i|X_i=x] - \lim_{x\uparrow{c}} \E[Y_i|X_i=x]. \]

Many methods are available for estimation, inference, and validation of RD designs within the continuity framework. The most common approach is to use (non-parametric) local polynomial methods to approximate the two regression functions near the cutoff. The implementation requires choosing a bandwidth around the cutoff, and then fitting two polynomial regressions of the outcome on the score---one above and the other below the cutoff---using only observations with scores within the chosen region around the cutoff as determined by the bandwidth. The bandwidth is typically chosen in an optimal and data-driven fashion to ensure transparency and to avoid specification searching. The most common strategy is to choose a bandwidth that minimizes the asymptotic mean squared error (MSE) of the RD point estimator \citep*[see][and references therein]{Calonico-Cattaneo-Farrell_2020_ECTJ}.

Although the implementation of local polynomial methods reduces to fitting two local, lower-order polynomials using least-squares methods, the interpretation and properties are different from the typical parametric least squares framework. Local polynomials are meant to provide a non-parametric approximation to the regression functions of the potential outcomes given the score. Because these functions are fundamentally unknown, the local polynomial approximation has an unavoidable misspecification error. When the bandwidth is chosen to be MSE-optimal, this approximation error appears in the standard distributional approximation and renders the conventional least squares confidence intervals invalid because conventional inference assumes that the approximation error is zero. \citet*{Calonico-Cattaneo-Titiunik_2014_ECMA} developed valid confidence intervals based on a novel distributional approximation for a test statistic in which the RD point estimator is adjusted using bias correction and also the standard error is modified to account for the additional variability introduced by the bias estimation step. These robust bias-corrected confidence intervals lead to valid inferences when using the MSE-optimal bandwidth, and remain valid for other bandwidth choices, in addition to having several other optimal properties \citep*{Calonico-Cattaneo-Farrell_2020_ECTJ}.

The canonical implementation of local polynomial estimation and inference in RD designs uses only the outcome and the score, without additional covariates. We discuss below how the local polynomial approach can be augmented to include pre-determined covariates in the two polynomial fits. In addition, we discuss how the local polynomial estimation and inference methods are adapted to incorporate covariates for other purposes such as heterogeneity analysis or extrapolation.

In the local randomization framework for RD analysis, the parameter of interest is defined differently because the core identifying assumptions are different from those in the continuity framework \citep*{Cattaneo-Frandsen-Titiunik_2015_JCI,Cattaneo-Titiunik-VazquezBare_2017_JPAM}. Because in the local randomization approach the central assumption is that the RD design creates conditions that resemble a randomized experiment in a neighborhood near the cutoff, the parameter is defined within this neighborhood---as opposed to at the cutoff as in the continuity approach. The local randomization parameter can be generically written as
\[\tau_\mathtt{SLR} = \frac{1}{N_\mathcal{W}} \sum_{i:X_i\in\mathcal{W}}\E_\mathcal{W}[Y_i(1) - Y_i(0)]\]
where $\mathcal{W}=[c-w,c+w]$ is a symmetric window around the cutoff, $N_\mathcal{W}$ is the number of units with $X_i \in \mathcal{W}$, $\E_\mathcal{W}$ denotes an expectation computed conditionally for all units with $X_i \in \mathcal{W}$, and two conditions are assumed to hold within $\mathcal{W}$: (i) the joint distribution of scores in $\mathcal{W}$ is known; and (ii) the potential outcomes are unaffected by the score $X_i$ in $\mathcal{W}$. Under these local randomization assumptions, estimation and inference can proceed using standard methods for the analysis of experiments, including Fisherian methods where the potential outcomes are assumed non-random and inferences are finite-sample exact, and Neyman or super-population methods where inferences are based on large-sample approximations with either non-random or random potential outcomes. In general, the parameter of interest can be written as  
\[\tau_\mathtt{SLR} = \frac{1}{N_\mathcal{W}} \sum_{i:X_i\in\mathcal{W}} \E_\mathcal{W}\Big[ \frac{T_i Y_i}{\P_\mathcal{W}[T_i=1]}\Big] 
                    - \frac{1}{N_\mathcal{W}} \sum_{i:X_i\in\mathcal{W}} \E_\mathcal{W}\Big[ \frac{(1-T_i) Y_i}{1-\P_\mathcal{W}[T_i=1]} \Big],
\]
where $\P_\mathcal{W}$ denotes a probability computed conditionally for all units  with $X_i \in \mathcal{W}$, which indicates a natural estimator by virtue of the known assignment mechanism within $\mathcal{W}$ and the other assumptions imposed.

The window $\mathcal{W}$ where the local randomization assumptions are assumed to hold is unknown in most applications. The recommended approach to choose this window in practice is based on balance tests performed on one or more pre-determined covariates \citep*{Cattaneo-Frandsen-Titiunik_2015_JCI}. The main idea behind this method is that the treatment effect on any pre-determined covariate is known to be zero, and thus a non-zero effect is evidence that the local randomization assumptions do not hold. Assuming that the covariate is correlated with the score outside of $\mathcal{W}$ but not inside, a test of the hypothesis that the mean (or other feature of the distribution) of the pre-determined covariate is the same in the treated and control groups should be rejected in all windows larger than $\mathcal{W}$, and not rejected in $\mathcal{W}$ or any window contained in it. The method thus consists of performing balance tests in nested (often symmetric) windows, starting with the largest possible window and continuing until the hypothesis of covariate balance fails to be rejected in one window and in all windows contained in it. 

The window $\mathcal{W}$ in the local randomization approach is analogous to the bandwidth in the continuity-based local polynomial analysis. An important distinction, however, is that the bandwidth can be chosen in a data-driven and optimal way using only data on the outcome and the score, while choosing the local randomization window $\mathcal{W}$ requires data on auxiliary pre-determined covariates. In this sense, the local polynomial analysis can be fully implemented without additional covariates, while the local randomization analysis typically cannot---unless $\mathcal{W}$ is known or chosen in an arbitrary manner. For further details on RD estimation and inference under the continuity-based and local randomization approaches, we refer the reader to \citet*{Cattaneo-Titiunik-VazquezBare_2020_BookCh} \citet*{Cattaneo-Idrobo-Titiunik_2020_CUP,Cattaneo-Idrobo-Titiunik_2022_CUP}, \citet*{Cattaneo-Keele-Titiunik_2022_MedReview}, \citet*{Cattaneo-Titiunik_2022_ARE}, and references therein.

\subsection{Efficiency and Power Improvements}

In the analysis of randomized experiments, it is well known that including pre-intervention covariates in the estimation of treatment effects can increase precision and statistical power, as discussed in standard textbooks on the analysis of experiments \citep{Rosenbaum2010-book,GerberGreen2012-book,ImbensRubin2015-book}. Covariate adjustment in experimental analysis can be implemented both before and after randomization has occurred. Adjustment before randomization is typically done via blocking or stratification, where units are first separated into groups according to their values of one or more pre-treatment covariates, and the treatment is then assigned randomly within groups. These methods are effective at increasing precision, but unfortunately they are not available for RD designs because the assignment of treatment is not under the control of the researcher, who typically receives the data after the treatment has already been assigned.\footnote{In recent years, there has been interest in designing ex-ante surveys for participants in RD designs. At the moment, the RD literature only has methodological developments for simple power calculation and survey sample design \citep*{Cattaneo-Titiunik-VazquezBare_2019_Stata}, but no work is available taking into account covariates for how to use ex-ante survey information to increase efficiency in the ex-post analysis.}

One popular ex-post method of covariate adjustment for experiments transforms the outcome to subtract from it a prior measure (also known as ``pre-test outcome'' or simply a ``pre-test'') that may come from a baseline survey or pilot study collected before the treatment was assigned. For example, instead of estimating the average treatment effect on the outcome, the researcher estimates the average treatment effect on the transformed outcome. The adjusted estimator is consistent for the average treatment effect, and leads to efficiency gains when the pre-test outcome is highly predictive of the post-treatment outcome. This idea can be easily implemented using modern non-parametric or machine learning methods.

Another popular method for ex-post covariate adjustment in experiments, and one of the most widely used, is to add pre-treatment covariates to a linear regression fit between the outcome and the treatment indicator. This regression-based covariate adjustment typically leads to efficiency gains when the added covariates are correlated with the outcome, and has been studied in detail in various contexts \citep*[see, for example,][and references therein]{ImbensRubin2015-book}. An interesting finding in this literature is that a regression model that incorporates the pre-intervention covariates and their interaction with the treatment variable is more efficient than a simple covariate adjusted estimator (i.e., excluding the interaction term), unless the treatment and control groups have equal number of observations or the covariates are uncorrelated with the individual treatment effect.

These ideas can be applied to the analysis of RD designs with suitable modifications. In the local randomization RD approach, the  particular properties of covariate adjustment follow directly from applying the results from the literature on experiments to the window around the cutoff where the local randomization assumptions are assumed to hold. Depending on the assumptions imposed (Fisherian, Neymann, or super-population), and the specific estimation and inference methods considered, the efficiency gains may be more or less important. 

Furthermore, regression-based covariate adjustment can also be used in the local randomization approach to relax some of the underlying assumptions. For example, \citet*{Cattaneo-Titiunik-VazquezBare_2017_JPAM} model the potential outcomes as polynomials of the running variable, to allow for a more flexible functional form of the potential outcomes inside the window $\mathcal{W}$ where the local randomization assumptions are assumed to hold. This model can be extended directly to include pre-determined covariates in the polynomial model, in addition to the running variable. 

In the continuity-based framework, applying the same arguments is not immediate because, although the local polynomial RD estimator is based on linear regressions above and below the cutoff, these regressions are non-parametric and the average treatment effect is estimated by extrapolating to the cutoff point. The properties of regression-based covariate adjustment in this context were first studied by \citet*{Calonico-Cattaneo-Farrell-Titiunik_2019_RESTAT}, who show that the inclusion of covariates other than the score in a local polynomial analysis can lead to asymptotic efficiency gains, if carefully implemented. The authors also show that covariate adjustment can result in unintended changes to the parameter that is being estimated depending on how the covariates are introduced in the estimation, a topic we revisit in Section \ref{sec: Can Covariates Fix a Broken RD Design?}. 

The standard local linear estimator of the RD treatment is implemented by running the weighted least squares regression of $Y_i$ on a constant, $T_i$, $X_i$, and $X_iT_i$ using only units with scores inside the chosen bandwidth, $X_i\in[c-h,c+h]$, and applying weights based on some kernel function. This leads to a point estimator, which can be interpreted as MSE-optimal if the bandwidth employed is chosen to minimize the mean squared error of the RD point estimator. This estimator includes only the score $X_i$, and thus it is said to be ``unadjusted'' because it does not incorporate any additional pre-treatment characteristics of the units. As discussed above, under standard continuity assumptions, the unadjusted local linear estimator $\hat{\tau}_{\mathtt{SRD}}$ is consistent for the continuity-based RD treatment effect $\tau_{\mathtt{SRD}}$, and robust bias-corrected inference can be employed as is now standard in the literature.

In practice, it is common for researchers to adjust the RD estimator by augmenting the local polynomial specification with the additional pre-determined covariates $Z_i$. \citet*{Calonico-Cattaneo-Farrell-Titiunik_2019_RESTAT} show that augmenting the local polynomial specification by adding covariates in an additive-separable, linear-in-parameters way that imposes the same common intercept for treated and control groups leads to a covariated-adjusted RD estimator that remains consistent for the canonical sharp RD treatment effect $\tau_{\mathtt{SRD}}$, while offering a reduction in its variance in large samples \citep*[see also][]{Ma-Yu_2022_wp}. The key required restriction, of course, is that the covariates are pre-intervention, and that they have non-zero correlation with the outcome. More specifically, whenever the effect of the additional pre-intervention covariates on the potential outcomes near the cutoff is (approximately) the same in the control and treatment groups, augmenting the specification with these covariates can lead to efficiency gains. Furthermore, the authors also show that other approaches commonly used in the applied RD literature for regression-based covariate adjustment can lead to inconsistent estimators of the parameter of interest $\tau_{\mathtt{SRD}}$.
 
More recently, \citet*{Arai-Otsu-Seo_2021_wp} proposed a high-dimensional implementation of the covariate-adjusted local polynomial regression approach of \citet*{Calonico-Cattaneo-Farrell-Titiunik_2019_RESTAT}, which allows for selecting a subset of pre-intervention covariates out of a potentially large covariate pool. Their method is based on penalization techniques (Lasso), and can lead to further efficiency improvements in practice. More generally, combining modern high-dimensional and machine learning methods can be a fruitful avenue for further efficiency improvements via covariate adjustment in RD settings. For example, it is straightforward to see that non-parametric or machine learning adjustments can be applied directly to the outcome variable to then employ the residualized outcomes in the subsequent local polynomial analysis, a procedure that can lead to further efficiency and power improvements. 

In sum, our discussion highlights several important issues. First, standard methods for covariate adjustment in the analysis of experiments can in principle be applied to RD designs with suitable modifications depending on the specific conceptual framework used. These methods and their properties apply immediately to RD estimation and inference in a local randomization framework, within the window $\mathcal{W}$ where the local randomization assumption is assumed to hold. The extension to continuity-based RD analysis is less immediate, as it requires considering the properties of the specific techniques at boundary points in a non-parametric local polynomial regression setup. However, there is a critical conceptual point: when using covariates for efficiency gains in RD designs, adjusting for those covariates should not change the RD point estimate. Regardless of the methods considered, covariate adjustment should not change the parameter of interest, and therefore unadjusted and adjusted point estimators should be similar in applications. 

Second, pre-intervention covariates can also affect tuning parameter selection and implementation of the RD design more broadly. For example, in the local randomization framework, covariates can be used to select the window where the key assumptions hold and to relax those assumptions, while in the continuity-based framework they can be used to further refine bandwidth selection for local polynomial estimation and robust bias-corrected inference.

Finally, while augmenting RD analysis with covariates to increase efficiency is a principled goal, researchers should be careful to avoid using covariate adjustment for specification searches. In empirical studies, researchers should always report unadjusted RD results first, and covariate-adjusted results second. Covariate adjustment can be implemented in many different ways, and different adjustments might lead to different results and conclusions. Instead of trying multiple ways of covariate adjustment at the time of analysis, researchers should pre-register their preferred adjustment method, and perform only that specification as opposed to trying multiple regression models. We return to this important point when discussing recommendations for practice in Section \ref{sec: Recommendations for Practice}.

\subsection{Auxiliary Information}

In the context of RD designs, covariate adjustment has also been proposed to address (i) missing data and measurement error, and (ii) to incorporate prior information via Bayesian methods.

A common concern in RD designs is that the RD running variable may be measured with error \citep*{Pei-Shen_2017_AIE}, and several approaches employing covariates have been recently proposed to address this issue. Generally speaking, key ideas from the imputation literature can be applied with appropriate modifications to RD designs. In the continuity-based framework, \citet*{BartalottiBrummetDieterle2021-JBES} study the case where the RD score is measured with error and this error changes across different groups of observations, and the researcher has access to auxiliary data that can be used to learn about the group-specific measurement error. The general procedure is to replace the mismeasured running variable by a ``corrected'' running variable of the same dimension that is estimated using the auxiliary covariates. Similarly, \citet*{DaveziesLeBarbanchon2017-JoE} study a fuzzy RD design where the running variable is measured with error, but the researcher has access to an additional covariate---the true running variable---that is observed for a sample of treated individuals. With this auxiliary data, the RD treatment effect at the cutoff can be identified. The procedure thus leads to heterogeneity tests that can be directly employed in empirical applications to assess whether the treatment affects different populations differently. Finally, in the local randomization framework, conventional methods can be used based on the standard missing-at-random assumption, where the missingness indicator and the potential outcomes are assumed to be independent conditional on pre-determined covariates. This assumption and associated methods can be applied directly within the window $\mathcal{W}$ to justify the use of imputation approaches.

The use of covariates has also been proposed to incorporate information via Bayesian approaches to RD analysis---which fall in a separate category, different from both local randomization and continuity-based RD approaches. \citet*{KarabatsosWalker2015-chapter} propose an infinite-mixture Bayesian model that allows the density of the outcome variable to depend flexibly on covariates. At a minimum, these covariates include the running variable and the treatment indicator, but other covariates can be added to the specification. A somewhat more general use of pre-determined covariates in Bayesian approaches to RD estimation is to use them to inform priors. Here, researchers sometimes use pre-treatment information rather than specific covariates. For example, in their study of the effect of prescribing statins to patients with high cardiovascular disease risk scores on their future LDL cholesterol levels, \citet*{GenelettiOKeeffeSharplesRichardsonBaio2015-StatinMed} use data on previous experimental studies  to formulate informative priors in a Bayesian approach to estimate RD effects. Finally, in a related approach, \citet*{ChibJacobi2016-JAppEcon} use principal stratification in a Bayesian framework to estimate complier average treatment effects in fuzzy RD designs. They model the imperfect compliance near the cutoff using a discrete confounder variable that captures different types of subjects: compliers, never-takers, and always-takers. Their setup leads to a mixture model averaged over the unknown subject types. The compliers' potential outcomes are modeled as a function of observed pre-determined covariates and the running variable, plus an idisyoncractic shock that follows a t-distribution. Given a prior distribution on the type probabilities, the posterior distribution of the complier average treatment effect is obtained by Markov chain Monte Carlo (MCMC) methods.

These types of methods have not been widely adopted in practice, but they provide useful examples of the different roles that covariate adjustment methods can have in RD applications. Importantly, in all these methods, the goal is ultimately to identify, estimate and conduct inference on the canonical (sharp and other) RD treatment effects, that is, without changing the parameter of interest. In the upcoming sections, we discuss covariate adjustment methods that do change the estimand, and therefore require careful interpretation.

\subsection{Treatment Effect Heterogeneity}
 
Another important use of covariates for RD analysis is to assess whether the treatment has different effects for different subgroups of units, where the subgroups are defined in terms of observed values of pre-determined covariates. The use of covariates to explore heterogeneity has a long tradition in the analysis of both experimental and non-experimental data, as researchers are frequently interested in assessing the effects of the treatment for different subpopulations. From a broader perspective, this analysis is equivalent to estimating treatment effects conditional on pre-treatment characteristics, which can typically be implemented using standard non-parametric or machine learning methods.

In RD designs, and regardless of the conceptual framework employed, the analysis of treatment effect heterogeneity necessarily changes the parameter of interest. For example, in the continuity-based framework, the parameter of interest is no longer $\tau_\mathtt{SRD}$, but rather
\[\tau_\texttt{CSRD}(z) = \E[Y_i(1)-Y_i(0)|X_i=c,Z_i=z],\]
where $Z_i$ denotes the additional pre-intervention covariate. The parameter $\tau_\texttt{CSRD}(z)$ corresponds to the (conditional, sharp) RD treatment effect at the cutoff for the subpopulation with covariates $Z_i=z$. Naturally, under appropriate assumptions, the canonical sharp RD treatment effect $\tau_\texttt{SRD}$ becomes a weighted average of the conditional RD treatment effects, with weights related to the conditional distribution of $Z_i|X_i$ near the cutoff. Similar ideas apply to the local randomization framework, depending on the specific assumptions invoked.

When the covariates $Z_i$ are few and discrete, the simplest strategy to explore heterogeneity is to conduct the RD analysis for each subgroup defined by $Z_i=z$, for each value $z$, separately. For example, in a medical experiment, researchers may be interested in separately estimating treatment effects for patients in different age groups, e.g. patients between 45 and 64 versus patients 65 and above. In such cases, separate analyses by subgroups have the advantage of being entirely non-parametric and involving no additional assumptions. One disadvantage is that the number of observations in each subgroup is lower than in the overall analysis, which can reduce statistical power. Whether this is a limitation will depend on the number of observations and other features of the data generating process in each particular application. Practically, this type of analysis can also be implemented by generating indicator variables for each category of the pre-intervention covariates, and then conducting estimation and inference using a fully saturated, interacted local polynomial regression model; if the indicators cover all possible subgroups, this strategy is equivalent to estimating effects separately by subgroup. This approach is also valid with respect to estimation and inference, and can be deployed in both the continuity-based and local randomization frameworks. However, if the covariates are multi-valued or continuous, using an interactive model imposes additional parametric assumptions, local to the cutoff, and the validity of inferences will depend on the validity of these assumptions.
 
If the number of covariates is large and researchers wish to explore heterogeneity in multiple dimensions, estimating separate average treatment effects by subgroups may be infeasible or impractical. In this case, researchers are usually interested in exploring heterogeneity in a large number of covariate partitions without specifying subgroups a priori. Standard non-parametric or machine learning methods can be useful in this situation, as they allow researchers to learn the relevant subgroups from the data. Machine learning methods in the continuity-based RD framework require modifications, and have begun to be explored only recently. In particular, \citet*{Reguly2021-wp} explores heterogeneity of the RD effect in subpopulations defined by levels of pre-determined covariates, creating a tree where each leaf contains an RD effect estimated on an independent sample. The approach assumes a parametric q-th order polynomial in each leaf of the tree; under this parametric assumption and additional regularity assumptions, the method leads to subgroup point estimates and standard errors that can be used for inference.

A different continuity-based approach to examine RD treatment effect heterogeneity on different subpopulations is proposed by  \citet*{HsuShen2019-JoE}. The authors impose stronger continuity conditions than in the standard RD design, requiring (in the sharp RD case) continuity of the expectation of the potential outcomes conditional on both the running variable and the additional covariates, and continuity of the conditional distribution of the additional covariates given the running variable. Under these conditions, they derive conditional average treatment effects given the covariates, and propose tests for three hypothesis: (i) that the treatment is beneficial for at least some subpopulations defined by values of the pre-determined covariates, (ii) that the  treatment has any impact on at least some subpopulations, and (iii) that the treatment effect is heterogeneous across all subpopulations. 

\cite{HsuShen2019-JoE}  define the null and alternative hypotheses by conditional moment inequalities given both the running variable and the additional covariates, and convert these conditional moment inequalities into an inﬁnite number of unconditional-on-covariates moment inequalities (that is, inequalities that are conditional only on the running variable and not on the additional covariates). Once the null and the alternative hypotheses are re-defined as instrumented moment conditions, these are estimated with local linear polynomials. Under regularity conditions, the asymptotic distribution of the local polynomial estimators of the moment conditions can then be used to derive the distribution of test statistics under the null hypothesis. In a follow-up paper, \citet*{HsuShen2021-JAppEcon} employ similar methods to develop a monotonicity test to assess whether a conditional local average treatment effect in a sharp RD design or a conditional local average treatment effect for compliers in a fuzzy RD design has a monotonic relationship with an observed pre-determined covariate---that is, the null hypothesis is that the conditional local average treatment effect, seen as a function of a covariate $z$, is non-decreasing in $z$.

Covariate adjustments for heterogeneity analysis within the local randomization framework may be more challenging due to limited sample sizes. Since the local randomization framework is generically more appropriate for very small windows $\mathcal{W}$ around the cutoff, and these windows typically contain only a few observations, conducting estimation and inference for subsets of observations according to $Z_i=z$ can be difficult in practice, leading to limited statistical power.

In sum, there are several important issues when using covariate adjustment for the analysis of heterogenous RD treatment effects. First, focusing on treatment effect heterogeneity requires that researchers redefine the parameter of interest. Second, estimating different RD treatment effects for different subgroups defined by covariate values can be done in a non-parametric way by estimating fully saturated models---or equivalently, by analyzing each sub-group separately. When the covariate dimensionality is too high and sub-group analysis is not possible, machine learning methods may offer an attractive strategy to discover relevant subgroups. Finally, augmented models that include interactions between the RD treatment and non-binary covariates require additional parametric assumptions for identification and inference.

\subsection{Other Parameters of Interest and Extrapolation}

Covariate adjustment has also been proposed in RD designs to identify additional parameters of interest. This includes cases where covariates are used to define new parameters for subpopulations, similar to the case of heterogeneity analysis discussed above, and cases where covariates are used to extrapolate treatment effects for score values far from the cutoff. 

In some RD designs, researchers have access to special covariates, and in exploring heterogeneity along those covariates, they redefine the parameter of interest. For example, it is common in practice to see treatments that follow the RD assignment rule but use different cutoffs for different subpopulations of units, that is, $T_i = \I(X_i \geq c_j)$ for $j=1,2, \ldots, J$. In this case, the cutoffs $c_1, c_2, \ldots, c_J$ naturally define subpopulations of units, where each subpopulation is defined by the cutoff that those units are exposed to. \citet*{Cattaneo-Keele-Titiunik-VazquezBare_2016_JOP} introduced the Multi-cutoff RD design, and investigated how multiple cutoffs can be used to provide more information about RD effects. In particular, the heterogeneity of the average treatment effect across cutoffs can be explored directly by performing RD analysis separately for each subgroup of units exposed to each cutoff value, as discussed in the prior section. The authors also discuss the role of the commonly used normalizating-and-pooling strategy, which leads to a causal interpretation of the canonical RD treatment effect as a weighted average of the cutoff-specific RD treatment effect under additional assumptions. See \citet*{onder2019heterogeneous} for an empirical application.

In some cases, researchers may prefer to use the information provided by the multiple cutoffs in combination with additional assumptions to define new parameters of interest. For example, \citet*{Bertanha2020-JoE} studies a Multi-Cutoff RD setup where each individual is exposed to a particular cutoff and a particular treatment dosage that changes as a function of that cutoff. Under additional assumptions, the author uses the changes in treatment dosage applied to individuals at various levels of the running variable to explore average treatment effects for counterfactual policies that set the treatment dosage to levels not observed in the data. \citet*{Cattaneo-Keele-Titiunik-VazquezBare_2021_JASA} combine a Multi-Cutoff RD design with additional assumptions to define the average treatment effect at a particular cutoff for a subpopulation of units exposed to a different cutoff value. Both approaches lead to extrapolation of RD treatment effects---that is, to parameters that capture the average effect of the treatment at values of the score different from the original cutoff to which units were exposed.
 
Another common setup involves an RD design observed in two periods of time, where the treatment of interest is confounded by an additional treatment in one period but not the other. \citet*{GrembiNanniciniTroiano2016-AEJ} call this setup a differences in discontinuities design. In the first period, there are two treatments that are assigned with the same RD rule, that is, the probability of receiving each treatment goes from zero to one at the same cutoff. In the second period, only one treatment changes at the cutoff, but the other treatment---which is the treatment of interest---is not active. The authors propose an estimator that takes the difference between the first-period RD effect and the second-period RD effect. Whether the limit of this estimator is the standard RD parameter will depend on the particular assumptions imposed. Under the usual parallel trends assumption of difference-in-differences designs, the estimand will remain unchanged. However, researchers could also make different assumptions about the change of the regression functions before and after, and redefine the parameters accordingly. 

When multiple cutoffs or time periods are not available, extrapolation of treatment effects can be done by using auxiliary pre-determined covariates. In this setting, with additional assumptions, researchers can learn about the treatment effect for units whose scores are not in the immediate neighborhood of the cutoff. For example, \citet{AngristRokkanen2015-JASA} propose a conditional independence assumption to study the effect of the treatment for observations far from the cutoff. This is formalized by assuming that, conditional on the covariates, the potential outcomes are mean-independent of the running variable. This ``selection on observables'' assumption (together with the standard common support assumption) immediately allows for extrapolation of the RD treatment effect, since it is assumed to hold for the entire support of the running variable. In this case, covariates allow for the identification of new parameters that capture the effect of the treatment at different values or regions of the running variable.

A similar conditional independence assumption has been used by in the context of a geographic or multi-score RD designs. In a geographic RD design, a treatment is assigned to units in a geographic area and withheld from units in an adjacent geographic area.  In this setup, units can be thought of as having a score (such as a latitude-longitude pair) that uniquely defines their geographic location and allows the researcher to calculate their distance to any  point on the boundary between the treated and control areas. The assignment of the treatment can then be viewed as a deterministic function of this score, and the probability of receiving treatment jumps discontinuously at the border that separates the treated and control areas. As discussed by \cite{KeeleTitiunik2015-PA}, seen in this way, the geographic RD design can be analyzed with the same continuity-based and local randomization tools of standard, one-dimensional RD designs. \citet*{KeeleTitiunikZubizarreta2015-JRSSA} consider a geographic RD application where units appear to choose their location on either side of the boundary in a  strategic way. They propose a conditional independence assumption according to which, for each point on the boundary, the potential outcomes are independent of the treatment assignment conditional on a set of covariates for units located in a neighborhood of that point \citep*[see also][for an extension to multiple dimensions]{diaz2020complex}. An important difference between their assumption and the conditional independence assumption proposed by \cite{AngristRokkanen2015-JASA} is that the former imposes conditional independence in a neighborhood of the cutoff and thus allows for extrapolation only within a neighborhood, while the latter imposes (mean) conditional independence along the entire support of the running variable and thus allows for extrapolation in the entire support of the running variable. Despite the differences, in both cases covariates are used for identification purposes, to define new parameters of interest that capture the average treatment effect at values of the running variable that are different from the cutoff.

Another use of auxiliary covariates for extrapolation purposes is proposed by \cite{WingCook2013-JPAM}, who augment the usual RD design with an exogenous 
measure of the outcome variable. Under the assumption that the exogenous outcome data can be used to consistently estimate the regression function of the actual outcome on the score in the absence of the treatment, treatment effects can be extrapolated to values of the score other than the actual cutoff used for assignment. 

In sum, additional covariates can be used in the RD design to define new parameters of interest, including parameters capturing treatment effects for units with score values different from the original cutoff. These covariates are sometimes part of the original RD design, as in the case of multiple cutoffs or multiple time periods, while in other settings they are external to the design and must be collected separately, as in the case of auxiliary unit characteristics or exogenous outcomes.

\section{Can Covariates Fix a Broken RD Design?}
\label{sec: Can Covariates Fix a Broken RD Design?}

In a randomized experiment, the fact that the treatment is randomly assigned implies that the treatment variable $D_i$ is independent of both potential outcomes $(Y_i(1),Y_i(0))$ and predetermined covariates $Z_i$. This means that, in experimental settings, we have independence between potential outcomes and treatment assignment, $(Y_i(1),Y_i(0)) \inde D_i$, and also conditional independence between potential outcomes and treatment assignment given the covariates, $(Y_i(1),Y_i(0)) \inde D_i | Z_i$. At the same time, it is well known that the assumption of conditional independence given covariates alone, $(Y_i(1),Y_i(0)) \inde D_i | Z_i$, does not imply $(Y_i(1),Y_i(0)) \inde D_i$. In observational studies, researchers often assume that $(Y_i(1),Y_i(0)) \inde D_i | Z_i$ in order to identify treatment effects. This assumption directly justifies covariate adjustment via regression models, inverse probability weighting, and other matching methods.

This observation has motivated some researchers to employ covariate adjustment to ``fix'' invalid RD designs where the identifying assumptions are suspected not to hold. In these cases, it is common to encounter RD analyses that include fixed effects or other covariates via additive linear adjustments in (local) polynomial regression estimation. In the methodological RD literature, adjustment for covariate imbalances has been proposed using multi-step non-parametric regression \citep*{FrolichHuber2019-JBES} and inverse probability weighting \citep*{Peng-Ning_2021_PMLR}.

In the continuity-based framework, covariate adjustment for imbalances in RD designs is done within a framework where $\E[Y_i(t)|X_i=x]$ is discontinuous at the cutoff but $\E[Y_i(t)|X_i=x, Z_i=z]$, seen as a function of $x$, is continuous in $x$, for $t=0,1$. These assumptions imply that the conditional distribution of $Z_i|X_i$ must be discontinuous at the cutoff, which in turn implies that canonical average RD treatment effects are necessarily not identifiable in general. Rather, in the absence of additional strong assumptions, the estimand that can be identified in this case is a weighted average of treatment effects where the weights depend on the conditional distributions of $Z_i|X_i$ for control and treatment units, which are different from each other \citep*{Calonico-Cattaneo-Farrell-Titiunik_2019_RESTAT,FrolichHuber2019-JBES}. As a consequence, in imbalanced RD designs, canonical RD treatment effects are not identifiable in general, and by implication covariate adjustment cannot fix a broken RD design. In the local randomization framework, the situation is analogous. 

In conclusion, for invalid RD designs where important covariates are imbalanced at the cutoff, covariate adjustment cannot restore identification of the canonical RD treatment effect. At best, covariate adjustment can identify some weighted average of the conditional treatment effects, $\E[Y_i(1)|X_i=x, Z_i=z]-\E[Y_i(0)|X_i=x, Z_i=z]$, with weights determined by properties of the two distinct conditional distributions of $Z_i|X_i$ below and above the cutoff. Unfortunately, in most applications, this covariate-adjusted estimand is not a parameter of interest and requires strong, additional assumptions on the underlying data generating process to deliver a useful parameter of interest. In other words, the main advantages of canonical RD designs in terms of identification are lost once the key identifying assumptions fail, and cannot be restored by simply using covariate adjustment methods without additional untestable assumptions.

\section{Recommendations for Practice}\label{sec: Recommendations for Practice}

When employed correctly, baseline covariates can be useful for the analysis and interpretation of RD designs. When pre-intervention covariates are available, they can be used for two main purposes without affecting the main RD identification strategy: (i) to improve efficiency and power, and (ii) to define new parameters of interest. However, analysts should keep in mind that canonical RD designs do not necessitate covariate adjustments, and therefore researchers should always report unadjusted RD treatment effect estimates and associated unadjusted robust bias-corrected inference methods.

Covariate adjustment for efficiency and power improvement is a natural way of incorporating baseline covariates in the RD analysis. This can be done by using regression adjustments, or via other well-known approaches from the experimental literature such as outcome adjustments, inverse probability weighting, or matching methods. Importantly, because the treatment effect of interest is the same with and without covariate adjustment, researchers should always report both unadjusted and covariate-adjusted RD results. The main and only goal of this practice is to improve efficiency and power (e.g., shorten confidence intervals), but the RD point estimate should remain unaffected, which should be borne out by the empirical analysis. Furthermore, it is important to guard against p-hacking in this context. Researchers should be cognizant of the dangers of multiple hypothesis testing when incorporating covariate adjustment in the RD analysis. One possible solution to address this concern is pre-registration of the analysis, which requires practitioners to declare ex-ante the covariate adjustment approach to be used in the subsequent RD analysis in other to avoid ex-post specification searching.

Pre-intervention covariates can also be employed for identification of other useful RD treatment effects. In Section~\ref{sec: Covariate Adjustment in RD Designs}, we discussed different examples such as treatment effect heterogeneity, multi-cutoff, geographic and multi-score designs, and extrapolation. These approaches offer relatively principled ways of incorporating covariates into an RD analysis. In the case of treatment effect heterogeneity, no substantial additional assumptions are needed, particularly in cases where the covariates are discrete and low-dimensional, in which case subset analysis is a natural way to proceed empirically. In the other cases, additional assumptions are needed in order to exploit covariate adjustment for identification of other meaningful RD treatment effects. These covariate adjustment practices are reasonable, and can be used in applications whenever the necessary additional assumptions are clearly stated and judged to be plausible. Again, it is important to guard against model specification searching, just like when employing pre-interventation covariates for efficiency and power improvements.  

Finally, an important misconception among some practitioners and methodologists is that covariates can be used to restore identification or ``fix'' an RD design where observations just above the cutoff are very different from observations just below it---in other words, a design where there is covariate imbalance at or near the cutoff and the RD assumptions are not supported by the empirical evidence. However, as we discussed in Section~\ref{sec: Can Covariates Fix a Broken RD Design?}, covariate adjustment in such cases requires strong additional assumptions, and cannot in general recover canonical RD treatment effects. In the absence of additional assumptions, adjusting for imbalanced covariates will at best recover other estimands at the cutoff that are unlikely to be of substantive interest in applications.

\section{Conclusion}\label{sec: Conclusion}

We provided a conceptual overview of the different approaches for covariate adjustment in RD designs. We discussed benign approaches based on pre-intervention (or pre-determined) covariates, including methods for efficiency and power improvements, heterogeneity analysis, and extrapolation. Those methods are principled and generally valid under additional reasonable assumptions on the data generating process. However, we also highlighted a natural tension between incorporating covariates in an RD analysis and issues related to p-hacking and specification searching. As a consequence, researchers employing covariate adjustment in RD designs should always be cognizant of issues related to model/covariate selection when implementing those methods in practice. Last but not least, we addressed the common misconception that covariate adjustment can fix an otherwise invalid RD design, and discussed how those methods can at best recover other estimands that may not be of interest. 

In conclusion, covariate adjustment can be a useful additional tool for the analysis and interpretation of RD design, when implemented carefully and in a principled way. In canonical RD settings it should never replace, only complement, the basic RD analysis based on the score and outcome variables alone. In other settings, covariate adjustment can offer additional insights in terms of heterogeneous or other treatment effects of interest, under additional design-specific assumptions.

%%%%%%%%%%%%%%%%%%%%%%%%%%%%%%%%%%%%%%%%%%%%%%%%%%%%%%%%%%%%%%
% References
%%%%%%%%%%%%%%%%%%%%%%%%%%%%%%%%%%%%%%%%%%%%%%%%%%%%%%%%%%%%%%

\clearpage
\onehalfspacing

\bibliographystyle{jasa}
\bibliography{CKT_2022_HandbookCh.bib}
\clearpage

\end{document}